\documentclass[prc,aps,preprint,showpacs]{revtex4}
\usepackage{graphicx}

\begin{document}
\title{Shell-structure of one-particle resonances in deformed potentials} 

\author{ Ikuko Hamamoto$^{1,2}$ }

\affiliation{
$^{1}$ {\it Riken Nishina Center, Wako, Saitama 351-0198, Japan } \\ 
$^{2}$ {\it Division of Mathematical Physics, Lund Institute of Technology 
at the University of Lund, Lund, Sweden} }   




\begin{abstract}
Shell structure of low-lying neutron resonant levels in axially-symmetric
quadrupole-deformed potentials is discussed, which seems analogous to that of
weakly-bound neutrons.  As numerical examples, nuclei slightly outside the
neutron-drip-line, $^{39}_{12}$Mg$_{27}$ and $^{21}_{6}$C$_{15}$, are studied. 
For the lowest resonance I obtain $I^{\pi}$ = $\Omega^{\pi}$ = 5/2$^{-}$ for 
$^{39}$Mg which is likely to be prolately deformed, while $I^{\pi}$ =
$\Omega^{\pi}$ = 1/2$^{+}$ may be assigned to the nucleus $^{21}$C which may be
oblately deformed.  Consequently, $^{21}$C will not be observed as a
recognizable resonant state, in agreement with the experimental information.    

\end{abstract}

\pacs{21.60.Ev, 21.10.Pc, 24.30.Gd, 27.30.+t, 27.40.+z}

\maketitle

\newpage

\section{INTRODUCTION} 
In nuclei towards neutron-drip-line a systematic change 
of neutron-shell-structure from the one well-known around the Fermi
level of stable nuclei  
has been discussed both theoretically and experimentally, especially in
connection with recent experimental data which have been obtained by using
radioactive ion beams.  
The possible disappearance of neutron magic numbers N=8 or 20 or 28 is 
an example.     
The systematic change can come from the fact that weakly-bound  
neutron orbits with larger orbital angular-momenta $\ell$ are sensitive to the 
strength of the potential, while those with smaller $\ell$ are much less
sensitive. 
This is because 
when eigen energies of bound one-neutron orbits  
approach zero, due to the strong $\ell$-dependence of the centrifugal potential 
the major part of the wave functions of orbits 
with smaller $\ell$ (thus, smaller $j$) lies outside the potential provided 
by well-bound core-nucleons, 
while that of orbits with larger $\ell$ (thus, larger
$j$) stays inside the potential.  
Consequently, for example, 
the energy of the 
weakly-bound $2p_{3/2}$-shell approaches that of 
the weakly-bound $1f_{7/2}$-shell. 
Namely, the magic number N=28 known around the Fermi level of 
stable nuclei disappears.  

Corresponding change of the shell-structure can be seen also in deformed nuclei,
though the characteristic behavior of orbits with smaller $\ell$ becomes 
somewhat vague because the wave functions of 
one-particle orbits in deformed potentials, in general, 
contain various $\ell$ components. An exceptionally unique case is neutron 
one-particle orbits with $\Omega^{\pi}$ = 1/2$^{+}$, where $\Omega$ expresses 
the component of one-particle angular-momentum along the symmetry axis of 
the axially-symmetric quadrupole-deformed potential.  In this case 
the $\ell$=0 component becomes dominant 
in deformed neutron wave-functions in the limit that 
the eigenenergy $\varepsilon_{\Omega}$ $(< 0)$ 
approaches zero, irrespective of the nature of orbits, because of the absence 
of the centrifugal barrier for the $\ell = 0$ channel.  In other words, 
in the limit the wave functions of one-neutron orbits 
with $\Omega^{\pi}$ = 1/2$^{+}$ in the  
deformed potential become $s_{1/2}$ wave-functions.    

While in the spherical limit an orbit with $\Omega^{\pi}$ = 1/2$^{+}$ 
can be created from every positive-parity
orbits such as $s_{1/2}$, $d_{3/2}$, $d_{5/2}$, $g_{7/2}$, $g_{9/2}$, ...,  
in deformed nuclei all one-particle orbits with 
$\Omega^{\pi}$ = 1/2$^{+}$ acquire
$\ell = 0$ components induced by the deformation.
Similarly, all one-particle orbits with $\Omega^{\pi}$ = 
1/2$^{-}$ in deformed potentials have some amount of $p_{1/2}$ component. 
 
In the case of deformed shape the component with the smallest orbital
angular-momentum $\ell_{min}$ 
in one-particle wave-functions plays a crucial role in the behavior of 
both weakly-bound and
resonant one-particle orbits.  Both the presence or absence of one-particle
resonance and the decay width can be determined decisively by the value of 
$\ell_{min}$, which is defined as the smallest orbital angular-momentum among
orbital angular-momenta of components in the wave function, 
and the magnitude of the $\ell$ = $\ell_{min}$ 
component in the wave function.    
It should be also noted that among an infinite number of positive-energy 
one-particle levels one-particle resonant levels are most important 
in many-particle correlations of bound nuclei.  
See Refs. \cite{IH07,IH12,IH16} for the characteristic 
feature of one-particle orbits unique in deformed potentials.

The above-mentioned shell-structure change has been discussed mostly 
for one-particle 
bound states, in connection with available experimental data on
nuclei close to but inside the drip line.    
However, since the component of one-particle wave-functions inside the nuclear
potential will contribute to the determination of one-particle resonant 
energies,
it is easily understood that the same kind of shell-structure  
as observed for weakly-bound neutrons is expected for
low-lying neutron resonant levels, as far as the resonances can be
well-defined with narrow decay-widths.    
Some numerical examples can be seen, for example, in Fig. 4 of Ref. \cite{IH07}
where the resonant energy ($\varepsilon > 0$) of the neutron $2p_{3/2}$ orbit 
is shown to 
approach that of the neutron $1f_{7/2}$ orbit in the way similar to the
energies of weakly-bound neutrons, or in
Fig. 3 of Ref. \cite{IH12}, which was presented in connection with the presence
(or absence) of the N=50 magic number for nuclei towards the neutron-drip-line. 
 
From my experience in both stable and drip-line nuclei I expect
that low-lying levels of odd-A deformed nuclei can well be analyzed  
in terms of one-particle motion in the deformed potential 
provided by the even-even core.   
For example, see the successful quantitative
analysis of various kinds of experimental data on odd-A nuclei in 
Ref. \cite{BM75}.  
Thus, in order to see the shell-structure change in low-lying neutron resonant
levels, I would examine odd-N nuclei, of which the neutron Fermi level lies
slightly above zero.      
It is emphasized that the shell-structure of one-particle resonant 
levels 
has no chance to be properly studied by conventional shell-model calculations,   
in which harmonic-oscillator wave-functions are used.  
  
In this respect, the nucleus $^{39}_{12}$Mg$_{27}$, which was verified to lie
outside the neutron-drip-line \cite{NO02}, is very interesting, because      
the neutron separation energy estimated from systematic trends of atomic mass 
\cite{MW12} is $-$130 keV. Thus, it is expected that the lowest resonance of 
the nucleus lies presumably
less than a few hundreds kev outside the drip line, and 
depending on the angular-momentum it may be observed as a sharp resonance.      
Even only the energies and widths of low-lying resonances of $^{39}$Mg will be 
sufficient to provide
the valuable information on the shell structure of one-particle resonant levels
for the possible deformed shape.  
If $^{39}$Mg is spherical and the shell-structure of neutron resonant levels  
is similar 
to something known in stable nuclei, one expects that the lowest resonance  
should be observed as a very sharp $f_{7/2}$ resonance presumably within a few 
hundreds keV in the continuum.  However, 
the proton number Z=12 is known to prefer prolate deformation as qualitatively 
understood from the presence of a large energy gap at Z=12 on the prolate side 
of one-proton energy diagrams as a function of quadrupole deformation 
(Nilsson diagrams). For example, see Fig. 3 of Ref. \cite{IH07}, noting that in
light nuclei the shell-structure of protons is nearly the same as that of
neutrons in
the case of well-bound orbits.   Thus, unless the neutron
number strongly prefers spherical shape, all even-even as well as odd-N 
neutron-rich Mg isotopes were predicted to be prolately deformed \cite{IH07}.  
Indeed, even-even neutron-rich Mg-isotopes with N=22-26 
towards the neutron drip-line are
found to be deformed by observing small values of $E(2_{1}^{+})$ and possible
$E(4_{1}^{+})/E(2_{1}^{+})$ ratios close to 3 \cite{PD13}.    
In Ref.\cite{HLC14} the direct two-proton removal from $^{42}$Si, which is
presumably oblately-deformed \cite{ST12, IH14}, is measured. 
The analysis of the inclusive
cross-section for the two-proton removal supports for the interpretation of a
prolately-deformed ground state of $^{40}$Mg \cite{HLC14}.  
The neutron separation energy of
$^{40}$Mg  is estimated  to be 1.740 MeV 
from the systematic trends \cite{MW12}.
On the other hand, in Ref. \cite{MT14} measured precise reaction cross sections
on C targets have been reported for $^{24-38}$Mg$_{12-26}$, from which deformed
shape of Mg-isotopes with N=20-26 is concluded.  In particular, 
the measured reaction cross section of $^{37}$Mg$_{25}$ is in good agreement 
with the interpretation of 
the $p$-wave halo nucleus, which is concluded 
from the Coulomb break-up reaction in
Ref. \cite{NK14}.    
From the available experimental information described above one may  
expect that $^{39}$Mg is prolately deformed and may be 
detected as a low-energy resonance.   
The spin-parity of the lowest resonance of $^{39}$Mg  
will be sensitive to the shell-structure
of one-particle resonant levels which will be analogous to that of 
weakly-bound neutrons.     
The case of $^{39}$Mg is interesting especially because there may be a chance 
to perform related experiments in a near future.       
I estimate a possible spin-parity of the lowest resonance 
of $^{39}$Mg, which will be 
identified as one-particle resonance  
with the energy of a few hundreds keV and a very narrow width.

Possible information on the unbound nucleus $^{21}_{6}$C$_{15}$ is of 
another
interests.  Using the single-proton removal reaction from a beam 
of $^{22}_{7}$N$_{15}$, the neutron-unbound nucleus $^{21}$C was searched for 
in Ref. \cite{SM13}.   Neutrons were detected in coincidence with $^{20}$C
fragments,
however, no evidence for low-lying states were found.  And, the reconstructed
[$^{20}$C + $n$]  decay-energy spectrum was reported to be described 
with an s-wave line shape.  
The neutron separation energy of $^{21}_{6}$C$_{15}$ estimated by using the
systematic trends of atomic mass \cite{MW12} is only  
$-$15 keV.  
The application of Woods-Saxon potentials to light nuclei 
such as C-isotopes may
be somewhat questioned, however, in the present article I examine the
shell-structure based on deformed Woods-Saxon potentials because my aim  
is mainly to study the one-particle resonant structure of $^{21}$C, the core
nucleus of which is $^{20}$C that may be oblately deformed.

While in Sec. II the model used and a brief summary of the definition of 
one-particle resonances in deformed potentials are 
given, in Sec.III practical examples of light nuclei are given.  In Sec. III.A  
the $pf$ shell structure in both bound and resonant 
one-particle levels, which are calculated by using deformed Woods-Saxon 
potentials with standard parameters (see Ref. \cite{BM69}), 
and the application of the idea to neutron-rich 
Mg isotopes, in particular to $^{39}$Mg, are presented.  
In Sec. III.B the $sd$ shell structure in both weakly-bound and resonant
one-particle levels in neutron-rich C-isotopes is described  
and, in particular, the interesting
case of the unbound nucleus $^{21}_{6}$C$_{15}$, which is expected to be only
slightly unbound but has not yet been observed, is presented.    
Conclusions and discussions are given in Sec IV.

\section{MODEL AND FORMULAS}

Since the definition of one-particle resonances in deformed potentials is 
described in detail in Ref. \cite{IH05}, 
a brief summary is given in the following.    
In the present article 
axially-symmetric quadrupole deformation is considered in the absence of pair
correlation.  
Writing the single-particle wave-function as 
\begin{equation}
\Psi_{\Omega} (\vec{r}) = \frac{1}{r} \sum_{\ell j} R_{\ell j \Omega} 
(r) \, {\bf Y}_{\ell j \Omega}(\hat{r}), 
\label{eq:twf}
\end{equation}
which satisfies 
\begin{equation}
H \, \Psi_{\Omega} = \varepsilon_{\Omega} \, \Psi_{\Omega}, 
\end{equation}
where $\Omega$ expresses the component of one-particle angular momentum 
along the symmetry axis, which is a good quantum number, and 
\begin{equation}
{\bf Y}_{\ell j \Omega}(\hat{r}) \equiv \sum_{m_{\ell}, m_{s}} C(\ell,
\frac{1}{2}, 
j;m_{\ell}, m_{s}, \Omega) \, Y_{\ell m_{\ell}}(\hat{r}) \, \chi_{m_{s}} .
\end{equation}
I study one-neutron resonances and use the 
Woods-Saxon potentials with standard parameters 
\cite{BM69} for the nuclear one-body potential.     
Then, the coupled equations for the radial wave-functions are written as 
\begin{equation}
\left(\frac{d^2}{dr^2} - \frac{\ell (\ell +1)}{r^2} + \frac{2m}{\hbar^2}( 
\varepsilon_{\Omega} - V(r) - V_{so}(r) ) \right) R_{\ell j \Omega}(r) = 
\frac{2m}{\hbar^2} 
\sum_{\ell^{'} j^{'}} \langle {\bf Y}_{\ell j \Omega} \mid V_{coupl} 
\mid {\bf Y}_{\ell^{'} j^{'}
\Omega} \rangle R_{\ell^{'} j^{'} \Omega}(r)
\label{eq:cpl}
\end{equation}
where 
\begin{eqnarray}
V(r) & = & V_{WS} \, f(r) \nonumber \\
f(r) & = & \frac{1}{1+exp \left(\frac{r-R}{a} \right)} \nonumber \\
k(r) & = & r V_{WS} \frac{df(r)}{dr} \nonumber \\
V_{coupl}  & = & - \beta \, k(r) \, Y_{20}(\hat{r}) \nonumber \\ 
\end{eqnarray}
For positive-energy ($\varepsilon_{\Omega} > 0$) one-particle levels  
I solve the coupled equations (\ref{eq:cpl}) 
in coordinate space for a given set of potential parameters, requiring 
\begin{equation}
R_{\ell j \Omega}(r) = 0   \qquad  \mbox{at} \quad r=0
\end{equation}
and the asymptotic behavior of $R_{\ell j \Omega}(r)$ for 
$r \rightarrow \infty$ as 
\begin{eqnarray}
R_{\ell j \Omega}(r) & \propto & cos(\delta_{\Omega}) \, \alpha_{c}r \,
j_{\ell}(\alpha_{c}r)- 
sin(\delta_{\Omega}) \, \alpha_{c}r \, n_{\ell}(\alpha_{c}r) \nonumber \\ 
& \rightarrow & sin(\alpha_{c}r \, + \, \delta_{\Omega} \, - \, \ell \frac{\pi}{2})
\label{eq:pbc}
\end{eqnarray}
where 
\begin{equation}
\alpha_c^2 \, \equiv \, \frac{2m}{\hbar^2} \varepsilon_{\Omega} 
\label{eq:a2}
\end{equation}
The quantity $\delta_{\Omega}$ in eq. (\ref{eq:pbc}) is called eigenphase.  

Solutions of Eq. (\ref{eq:cpl}) satisfying the above boundary conditions can 
be found for any $\varepsilon_{\Omega} > 0$ values, since the problem here 
is not an eigenvalue problem such as the problem of finding bound states.
It is noted that the eigenphase is common to all components of 
open channels, ($\ell, j$), of a given solution with $\Omega$.  For given
$\varepsilon_{\Omega}$ and potential one may obtain several solutions for
$\delta_{\Omega}$.  The number of solutions is indeed equal to 
that of wave-function components with different $(\ell, j)$.       

One-particle resonance in deformed potentials is obtained, when an eigenphase
$\delta_{\Omega}$ increases through $\pi / 2$ as $\varepsilon_{\Omega}$
increases.  In the limit of $\beta \rightarrow 0$ 
this definition, of course, agrees with the definition of one-particle
resonance in spherical potentials, which can be found in various text books.  

When I find one-particle resonant levels thus defined, I calculate
the intrinsic width of the resonance using the formula
\begin{equation}
\Gamma_{intr} \equiv \frac{2}{{\frac{d \delta_{\Omega}}{d
\varepsilon_{\Omega}}}|_{\varepsilon_{\Omega} = \varepsilon_{res}}}
\label{eq:width}
\end{equation}
where the denominator is calculated at the resonance energy.  
The value of $\Gamma_{intr}$ is usually determined decisively 
by the $\ell_{min}$ component of the wave function.  
In the case of band-head states, of which the spin-parity is 
$I^{\pi}$ = $\Omega^{\pi}$ with possible exceptions for $\Omega$ = 1/2,  
it is expected that the width ($\Gamma$) observed in the
laboratory is approximated by the above intrinsic width.  
This expectation is based on the following approximate picture: 
The structure of the band-head state may be generally little affected by
rotational perturbation (namely little Coriolis perturbation).   In order to
obtain the band-head state in the laboratory system, which is an eigenstate of
angular-momentum, various direction of the deformed intrinsic state must be
properly superposed (a kind of zero-point fluctuation related to rotational
motion).  Now, since the decay by neutron emission will occur in general 
much faster than
slow rotational motion, in the first approximation the process of 
neutron emission may be treated for a fixed direction of deformation (or the
deformed intrinsic state).  That means, the decay-width of neutron emission may
be estimated in the intrinsic system.  If so, the width $\Gamma$ to be observed
in the laboratory system is approximated by $\Gamma_{intr}$.   

As no one-neutron resonance with $s_{1/2}$ is obtained in spherical potentials 
because of the absence of centrifugal potential for the $\ell = 0$ channel, 
one-neutron resonance with  
$\Omega^{\pi}$ = 1/2$^{+}$  in deformed potentials is rare to be found. 
For example, see Fig.1 of Ref. \cite{IH05}.
However, even for the positive energy of more than several MeV 
the $\Omega^{\pi}$ = 1/2$^{+}$ 
one-neutron resonance can be present in
the case that the absolutely overwhelming components of the
wave function have larger $\ell$ values.  In such cases the one-neutron
resonance can exist because of the components with large $\ell$ 
while the decay width is controlled by the tiny $\ell$ = 0 component.   
For larger deformations one-neutron wave
functions usually contain an appreciable amount of $s_{1/2}$ component. 
As soon as the wave function starts to
contain an appreciable amount of $s_{1/2}$ component, 
the possible one-neutron resonant
state immediately decays via the $s_{1/2}$ channel and, then,  
the related eigenphase may not reach $\pi / 2$.     

Similarly, in spherical potentials with parameters appropriate for 
neutron-rich Mg isotopes 
one-neutron resonance with $p_{1/2}$ or $p_{3/2}$ may not be
obtained at the energy higher than, say about 1 MeV, due to the low centrifugal
barrier.  For
larger deformation of those potentials it is not often to obtain higher-lying 
one-neutron resonances with $\Omega^{\pi}$ = 1/2$^{-}$ or 3/2$^{-}$.
Only when the overwhelming 
components of the resonances have larger $\ell$ values, 
one-neutron resonances with $\Omega^{\pi}$ = 1/2$^{-}$ or 3/2$^{-}$ can survive 
even for larger positive energies.

In contrast, for example, 
in the $pf$-shell nuclei one-neutron resonances with $\Omega^{\pi}$ =
5/2$^{-}$ or 7/2$^{-}$ can be easily obtained up till positive energy 
of several MeV with
a meaningful amount of decay width, 
because the $\ell_{min}$ value of the components of the wave functions 
is 3, 
for which a pretty high centrifugal barrier is present.  

In the present work I take the rule that in odd-N nuclei the spin-parity 
of the band-head state 
of the low-lying band with $\Omega^{\pi}$  
is $I^{\pi}$ =
$\Omega^{\pi}$, where $\Omega$ expresses the angular-momentum component 
of the N-th neutron orbit along the symmetry axis for an appropriate quadrupole
deformation, except for the case of $\Omega$ = 1/2.
In the case of $\Omega$ = 1/2 the decoupling parameter, which is the diagonal
element of the particle-rotation coupling that can be calculated by using the
one-particle wave-function of the N-th neutron orbit, should be taken into
account.  
Examining experimental data there are no exceptions, to my knowledge, 
to the above rule in
the case of well-deformed odd-A nuclei.  
For typical examples, see the analysis of
$^{25}_{12}$Mg$_{13}$ on p.284 of Ref. \cite{BM75} or $^{175}_{70}$Yb$_{105}$ on
p.259 of Ref. \cite{BM75}.  
The picture of one-particle motion in deformed potentials works for deformed
nuclei indeed much better than that of one-particle motion in spherical
potentials for spherical nuclei.  This is mainly because the major part of the
($Y_{20}^{\ast} \cdot Y_{20}$) channel of the two-body quadrupole-quadrupole
interaction is absorbed into the quadrupole mean-field.

\section{PRACTICAL EXAMPLES}

\subsection{$pf$ shell and Mg-isotopes}
In Fig. 1 energies of one-neutron bound- and resonant-levels for the system
[$^{38}_{12}$Mg$_{26}$ + one neutron] are plotted as a
function of axially-symmetric quadrupole deformation parameter $\beta$, 
while in Fig. 2 energies of
one-neutron bound-levels for the system [$^{48}_{22}$Ti$_{26}$ + one neutron] 
are shown.  The nucleus $^{39}_{12}$Mg$_{27}$ is expected to lie 
outside the neutron-drip-line only by a few hundreds keV, 
while the nucleus $^{49}_{22}$Ti$_{27}$ is a typical stable odd-N $pf$-shell 
nucleus, of which the neutron separation energy S(n) is equal to  8.142 MeV, 
and the observed spin-parity of 
the ground state is 7/2$^{-}$.   
In Fig. 2 one-neutron resonant energies are
not plotted since they are not important for the present discussion.  
  
The parameters of Woods-Saxon potentials used in Figs. 1 and 2 are taken from
the set of standard parameters \cite{BM69} except for the depth in Fig. 1.   
The value $V_{WS}$ = $-$37.5 MeV used in Fig. 1 is shallower than 
the standard value for the proton and neutron numbers of $^{38}$Mg by 1.3
MeV, but it was chosen so that $^{39}$Mg is obtained as one-particle resonant
levels at a few hundreds keV.    
  
Examining Fig. 5 of Ref. \cite{IH07}, Fig. 2 of Ref. \cite{IH12} and Fig.1
of the present article it is clearly seen that when one-neutron  $1f_{7/2}$ and 
$2p_{3/2}$ states become weakly bound or one-particle resonant states they
become almost degenerate for spherical shape.  
The magic number N=28 disappears and, as a result of it,  
the nuclei with $N$ =21-28 may be deformed if the proton number allows the
deformation \cite{IH07}.  
Following the shell-structure change for spherical shape, 
the shell-structure of $pf$-shell 
in deformed shape changes as seen in the comparison 
between Fig. 1 and Fig. 2.  
In contrast, using the present Woods-Saxon potential with the standard
parameters the neutron number N=28 is a good magic number, of course,  
for spherical shape in stable nuclei, as seen in Fig. 2.     

The ground-state spin-parity of $^{33}_{12}$Mg$_{21}$ is known to be 3/2$^{-}$
\cite{DY07},
but that of neutron-rich Mg isotopes with N = 23 and 25 is not yet
experimentally pinned down.  
Studying the level structure of the prolate side of Fig. 1 one obtains
that the one-particle orbit occupied by the last odd-neutron
(namely the 21st neutron) of $^{33}$Mg 
is [330 1/2], $\Omega^{\pi}$ = 1/2$^{-}$ for $0 < \beta < 0.31$,  
[202 3/2], $\Omega^{\pi}$ = 3/2$^{+}$ for $0.31 < \beta < 0.42$ and 
[321 3/2], $\Omega^{\pi}$ = 3/2$^{-}$ for $0.42 < \beta < 0.55$. 
In the case of the
lowest-lying $\Omega^{\pi}$ = 1/2$^{-}$ orbit in the deformed $pf$ shell, 
namely the [330 1/2] orbit, 
which may be assigned to the ground-state configuration 
of $^{33}_{12}$Mg$_{21}$,  
the spin-parity of the lowest state (namely, the band-head state) 
to be observed is $I^{\pi}$ = 3/2$^{-}$ due to the
decoupling parameter coming from the main components $1f_{7/2}$ and $2p_{3/2}$
of the wave-function. 
Thus, the measured spin-parity $I^{\pi}$ = 3/2$^{-}$ of the ground state of 
$^{33}$Mg$_{21}$ can be naturally obtained for either $0 < \beta < 0.31$ or 
$0.42 < \beta < 0.55$ in Fig. 1.  
Furthermore, the calculated magnetic moment for the band-head state 
$I^{\pi}$ = 3/2$^{-}$ from the intrinsic [330 1/2] state is $-$0.88 $\mu_{N}$
for $\beta$ = 0.25 and that for the band-head state $I^{\pi}$ = 3/2$^{-}$ 
from the [321 3/2] state is $-$0.40 $\mu_{N}$ for $\beta$ = 0.50, while that for
the band-head state $I^{\pi}$ = 3/2$^{+}$ from the intrinsic [202 3/2] state is 
$+$0.91 $\mu_{N}$ for $\beta$ = 0.35, in comparison with the measured value of 
$-$0.7456(5) $\mu_{N}$ \cite{DY07}.  
  
The deformed 
$p$-wave halo nature of the ground state of $^{37}_{12}$Mg$_{25}$ observed
in \cite{NK14} is consistent with the neutron orbit 
with $\Omega^{\pi}$ = 1/2$^{-}$, which is the second-lowest 
$\Omega^{\pi}$ = 1/2$^{-}$ orbit on the prolate side of the $pf$-shell 
(see Fig. 5 of Ref. \cite{IH07}.) 

From the above consideration, 
it may be acceptable to assume that the 27th odd neutron in 
$^{39}$Mg$_{27}$
occupies the orbit with $\Omega^{\pi}$ = 5/2$^{-}$, which is denoted by
a long-dashed curve in Fig. 1 having the
positive energy of several hundreds keV.  From the present consideration 
one almost uniquely ends up with $I^{\pi}$ = $\Omega^{\pi}$ = 5/2$^{-}$ 
for the spin-parity of the lowest resonance of $^{39}$Mg, if
the nucleus is prolately deformed and the two one-particle levels, 
$1f_{7/2}$ and $2p_{3/2}$ at $\beta = 0$, are almost degenerate.   
Furthermore, on the prolate side 
the calculated one-particle energy of the $\Omega^{\pi}$ = 5/2$^{-}$ orbit 
originating from the $1f_{7/2}$ shell at $\beta$=0 is almost independent of 
the $\beta$ value.  
The almost constant (intrinsic) energy of the $\Omega^{\pi}$ = 5/2$^{-}$ level 
as a function
of $\beta$ seen in Fig. 1  
does not come from the property of weakly-bound level 
or low-lying resonant level, 
but it    
comes from the fact that the main component of
the level is $1f_{7/2}$ with a small admixture of $1f_{5/2}$.  

The calculated one-particle decay width of the I$^{\pi}$ = 5/2$^{-}$ state is 
0.01 MeV at 
$\beta \approx 0$ for the $1f_{7/2}$ resonant state 
at $\varepsilon_{res}$ = 0.32 MeV, 
and the width does not considerably change as the system
becomes prolately deformed, because the one-particle resonant energy is almost
constant and the main component mixed due to the deformation 
is $1f_{5/2}$, which has the same $\ell = 3$ orbital 
angular-momentum as that of $1f_{7/2}$. 
For example, the calculated intrinsic width $\Gamma_{intr}$ ($\approx \Gamma$)  
is 0.02 MeV at 
$\beta$=0.4 for the resonant $\Omega^{\pi}$ = 5/2$^{-}$ state at 
$\varepsilon_{res}$ = 0.62 MeV.    
If the positive energy of 
the lowest resonance of 
$^{39}$Mg deviates considerably from several hundreds keV, 
the width will be accordingly different.  

By the way, in the present example of $^{39}$Mg the one-particle orbit closest 
to the N = 27th $\Omega^{\pi}$ = 5/2$^{-}$ orbit on the prolate side is the
$\Omega^{\pi}$ = 1/2$^{-}$ orbit, the energy of which is denoted by a solid
curve in Fig. 1.  The particle-rotation coupling between the two states having
the particle orbits with
$\Omega^{\pi}$ = 5/2$^{-}$ and $\Omega^{\pi}$ = 1/2$^{-}$ is expected to be
pretty weak because of $\Delta \Omega$ = 2.

\subsection{$sd$ shell and C-isotopes}
The neutron-rich C-isotopes, in particular the nucleus $^{21}_{6}$C$_{15}$ lying
slightly outside the neutron-drip-line, are another interesting examples in the
sense different from Mg-isotopes.  The observed energies of 
the 2$_{1}^{+}$ state in $^{16}$C, $^{18}$C and $^{20}$C are 1.76, 1.58 and 1.61
MeV, respectively, while corresponding measured 
$B(E2; 2_{1}^{+} \rightarrow
0_{gs}^{+})$ values are 4.21$^{+1.26}_{-0.50}$ \cite{MP12}, 
3.64$^{+0.55}_{-0.61}$ 
\cite{PV12} and 7.5$^{+4.0}_{-2.1}$ e$^{2}$ fm$^{4}$ \cite{MP11}, respectively. 
In all those C-isotopes the main components of the 2$_{1}^{+}$ state are
expected to be neutron configurations, and it is not straightforward to obtain
the information on the shape from those B(E2)-values.  For example, 
in the shell-model calculations those B(E2)-values depend totally 
on the neutron E2 effective charge used.   The measured 
energies of those 2$_{1}^{+}$ states may be possibly interpreted as indicating 
deformed shape.   
For reference, the values of quadrupole deformation parameter $\beta$ 
for the ground states of 
$^{16}$C, $^{18}$C and $^{20}$C 
obtained in the Skyrme
Hartree-Fock (HF) calculation in Ref. \cite{HS04} are 0.34, 
0.36 and $-$0.35, respectively.     

The observed ground-state spin-parities of $^{17}_{6}$C$_{11}$ and
$^{19}_{6}$C$_{13}$ are 3/2$^{+}$ and 1/2$^{+}$, respectively, while the measured
neutron separation energies $S(n)$ of $^{17}$C and $^{19}$C are 
0.73 and 0.58 MeV, respectively.
The observed spin-parities of the ground states are in fact in agreement with
$I^{\pi}$ = $\Omega^{\pi}$ = 3/2$^{+}$ as the band-head state from the intrinsic
[211 3/2] state for
$^{17}_{6}$C$_{11}$ and $I^{\pi}$ = $\Omega^{\pi}$ = 1/2$^{+}$ 
as the band-head state from the intrinsic [211 1/2] 
state for $^{19}_{6}$C$_{13}$, both on the prolate side.  Namely, the
observed spin-parities of $^{17}$C and $^{19}$C are 
easily obtained if I assume
the prolate shape for those nuclei.  Furthermore, the measured magnetic moment
of $^{17}$C, $|\mu_{obs}|$ = 0.758 $\mu_{N}$ \cite{HO02} is in agreement with
the 
calculated value, $\mu_{calc}$ = $-$0.75 $\mu_{N}$, 
for the band-head state from
the intrinsic [211 3/2] neutron 
state at $\beta \approx 0.4$.
For reference, the values of spin-parity and quadrupole deformation of the
ground states of $^{17}$C and $^{19}$C, which are obtained in the HF calculation
of Ref. \cite{HS04}, are 
(3/2$^{+}$, $\beta$=0.39) and (3/2$^{+}$, $\beta=-$0.36), respectively.   

In Fig. 3 energies of one-neutron bound- and resonant-levels for the system
[$^{20}_{6}$C$_{14}$ + one neutron] are plotted as a function of $\beta$.  
The parameters of the Woods-Saxon potential are the standard parameters
\cite{BM69} except for the depth.  The used value $V_{WS}$ = $-$36.5 MeV 
is shallower than the standard value for the proton and neutron numbers of 
$^{20}$C by 1.3 MeV, but it was chosen so that the 15th neutron in $^{21}$C may
occupy a one-particle orbit with a positive energy 
when the nucleus is deformed.  
One may compare Fig. 3 with Fig. 1 of Ref. \cite{IH12}, which is drawn 
for the [$^{18}_{6}$C$_{12}$ + one
neutron] system where the standard parameters for the proton and
neutron numbers of $^{18}$C are used also for the value of
$V_{WS}$.

It is pointed out in Ref. \cite{IH14} that the particle-numbers 6 and 14 are 
magic numbers in the deformed harmonic-oscillator (ho) potential with the
oblate deformations ($\omega_{\perp} : \omega_3$) = (1:2) and (2:3).  Moreover,
also for Woods-Saxon potentials with standard parameters large energy gaps are
obtained at the particle numbers 6 and 14 for $\beta < -0.2$ .  See Fig.2 of
Ref. \cite{IH14}.   That means, the nucleus $^{20}$C has 
a very good chance of being
oblately deformed, while only a few even-even nuclei are known 
to be oblate in
the region of lighter nuclei with $6 \leq Z \leq 40$. 

From Fig. 3 it is seen that on the oblate side the 15th neutron of $^{21}$C 
will
occupy the $\Omega^{\pi}$ = 1/2$^{+}$ orbit, the lowest state of which is 
$I^{\pi}$ = 1/2$^{+}$.  The state cannot be obtained 
as a one-particle resonant state for $\beta < -0.1$,  
namely, the state has a very large width and can be hardly identified as a
state.    

In short, from the present consideration it is very likely that 
the unbound nucleus 
$^{21}$C is not observed as a resonant state 
though the possible lowest resonance may be 
expected to be only slightly outside the neutron-drip-line.  This result, 
in particular, the oblate shape of $^{21}$C, 
thereby the spin-parity $I^{\pi}$ = $\Omega^{\pi}$ = 1/2$^{+}$ for the lowest
state,  
is in good agreement with the observed fact \cite{SM13} that no evidence 
for a low-lying state has been yet found.  
Though ''no evidence'' in Ref. \cite{SM13} 
is consistent with the spin-parity $I^{\pi}$ = 1/2$^{+}$
coming from the interpretation that the 15th neutron is placed 
on the $2s_{1/2}$ orbit in the continuum assuming spherical shape, 
I want to point out that $^{21}$C is likely to be
oblately deformed and ''no evidence'' is consistent also with the oblate shape.

\section{Conclusions and discussions}   
I have shown that the spin-parity of the lowest resonance of 
$^{39}$Mg, which lies outside the neutron-drip line by a few hundreds keV, 
can be 5/2$^{-}$ and is expected to be a very sharp resonant state.   
Neutron-rich Mg isotopes with N $\geq$ 21 were expected and seem to be 
(prolately) deformed.
The spectroscopic study on the deformed odd-N isotopes are known to be
valuable for obtaining the information on the shell-structure of 
neutrons in deformed one-body potentials.  
It is expected that the shell-structure of low-lying neutron 
resonant levels is similar to that of weakly-bound neutron 
levels.  In particular, when the neutron $1f_{7/2}$ resonant level 
occurs at a few hundreds keV the neutron $2p_{3/2}$ resonant level will be 
found at nearly the same energy, though one-particle decay widths of
those two resonances can be different by two orders of magnitudes.  
The spectroscopic information on the
nucleus $^{39}$Mg$_{27}$ is very helpful for confirming the shell-structure of
one-particle resonant levels.  
I am looking forward to obtaining any experimental information on $^{39}$Mg.  

The nucleus $^{21}_{6}$C$_{15}$ is another interesting nucleus from 
the present point of
view.  The core nucleus $^{20}$C is expected to be oblately deformed based on 
good reasons.  If so, the system of [$^{20}$C + one neutron], namely
$^{21}_{6}$C$_{15}$, may be oblate.  Then, it is expected 
that though the nucleus 
$^{21}$C may lie only very slightly outside the neutron-drip-line, 
it will not be found as a recognizable resonant state, in
agreement with available experimental findings.
 
In the present work the pair-correlation is neglected since only light nuclei
are considered.  Even when the pair-correlation is effective in the present
nuclei, the spin-parity estimated for the lowest resonance of $^{39}$Mg and
$^{21}$C will not be changed while the estimated decay width may be changed.  

The author is thankful to Drs. P. Fallon and H. L. Crewford for the
communications about the possibility of performing experiments to obtain the
information on $^{39}$Mg.

\newpage

\noindent
{\bf\large Figure captions}\\
\begin{description}
\item[{\rm Figure 1 :}]
Calculated one-particle energies for neutrons in the potential given by the
core nucleus $^{38}_{12}$Mg$_{26}$ as a function of axially-symmetric quadrupole deformation.
Bound one-particle energies at $\beta$ = 0 are $-$5.90 and $-$4.27 MeV
for the $2s_{1/2}$ and $1_{d3/2}$ levels 
respectively, while
one-particle resonant $1f_{7/2}$ and $2p_{3/2}$ levels are obtained 
at +0.32  and +0.47 MeV (denoted by filled circles) 
with the width of 0.01 MeV and
+1.02 MeV, respectively.   
The $2p_{1/2}$ level is not obtained as one-neutron resonant level 
and, thus, is not
plotted.  The $1f_{5/2}$ resonant state is obtained at +5.83 MeV with the width
of 3.00 MeV and is denoted by a filled circle.  
The $\Omega^{\pi}$ = 3/2$^{-}$ level originating from the resonant 
$2p_{3/2}$ level at $\beta = 0$, which steeply bends upward for both
$\beta > 0$ and $\beta < 0$, is not  
obtained as a one-neutron resonant level for appreciable values of 
$\mid \beta \mid \neq 0$ and, thus, 
does not appear in the figure. 
Due to the same reason the $\Omega^{\pi}$ = 1/2$^{-}$ level on the oblate side,
which is connected to the resonant $2p_{3/2}$ level at $\beta$=0, cannot be
plotted.   
The neutron numbers, 20 and 24, which
are obtained by filling all lower-lying levels, are indicated with circles, for
reference.     
For simplicity, calculated widths of one-particle resonant levels are not shown.
The parity of levels can be seen from
the $\ell$-values denoted at $\beta$ = 0; $\pi$ = $(-1)^{\ell}$.

\end{description}

\begin{description}
\item[{\rm Figure 2 :}]
Calculated one-particle energies for neutrons in the potential given by 
the core nucleus $^{48}_{22}$Ti$_{26}$ 
as a function of axially-symmetric quadrupole deformation.
Bound one-particle energies at $\beta$ = 0 are $-$15.078, $-$10.01, $-$6.24, 
$-$4.05 and 
$-$2.86 and $-$0.22 MeV for the $1d_{3/2}$, $1f_{7/2}$, $2p_{3/2}$, $2p_{1/2}$, 
$1f_{5/2}$, and $1g_{9/2}$ levels, respectively, 
The neutron numbers, 24, 28 and 36, which are obtained 
by filling all lower-lying
levels, are indicated with circles. 
Since the Fermi level of the nucleus $^{49}$Ti lies around $-$8 MeV, one-neutron
resonant levels are not important in the present discussion and, thus, are not
shown in the figure.

\end{description}

\begin{description}
\item[{\rm Figure 3 :}]
Calculated one-particle energies for neutrons in the potential given by the core
nucleus $^{20}_{6}$C$_{14}$ as a function of axially-symmetric quadrupole
deformation.  Bound one-particle energies at $\beta = 0$ are $-$6.93, $-0.67$,
and $-0.44$ MeV for the $1p_{1/2}$, $1d_{5/2}$ and $2s_{1/2}$ levels,
respectively.  One-particle resonant $1d_{3/2}$ level is obtained at 
$+$4.13 MeV with the width of 4.40 MeV and is denoted by a filled circle.  
For simplicity, calculated widths of one-particle resonant levels are not 
shown.  The $\Omega^{\pi}$ = 1/2$^{+}$ level connected to the bound 
$2s_{1/2}$ level at $\beta$=0 cannot be obtained as a one-particle resonant 
level either for
$\beta < -0.1$ MeV or $\beta > 0.3$ MeV.   
The $\Omega^{\pi}$ = 1/2$^{+}$ resonant level connected to the 
$1d_{3/2}$ state at 
$\beta$=0 is not obtained for $\beta > 0.46$ as a one-particle resonant level 
because of the increasing $\ell = 0$ component.

\end{description}

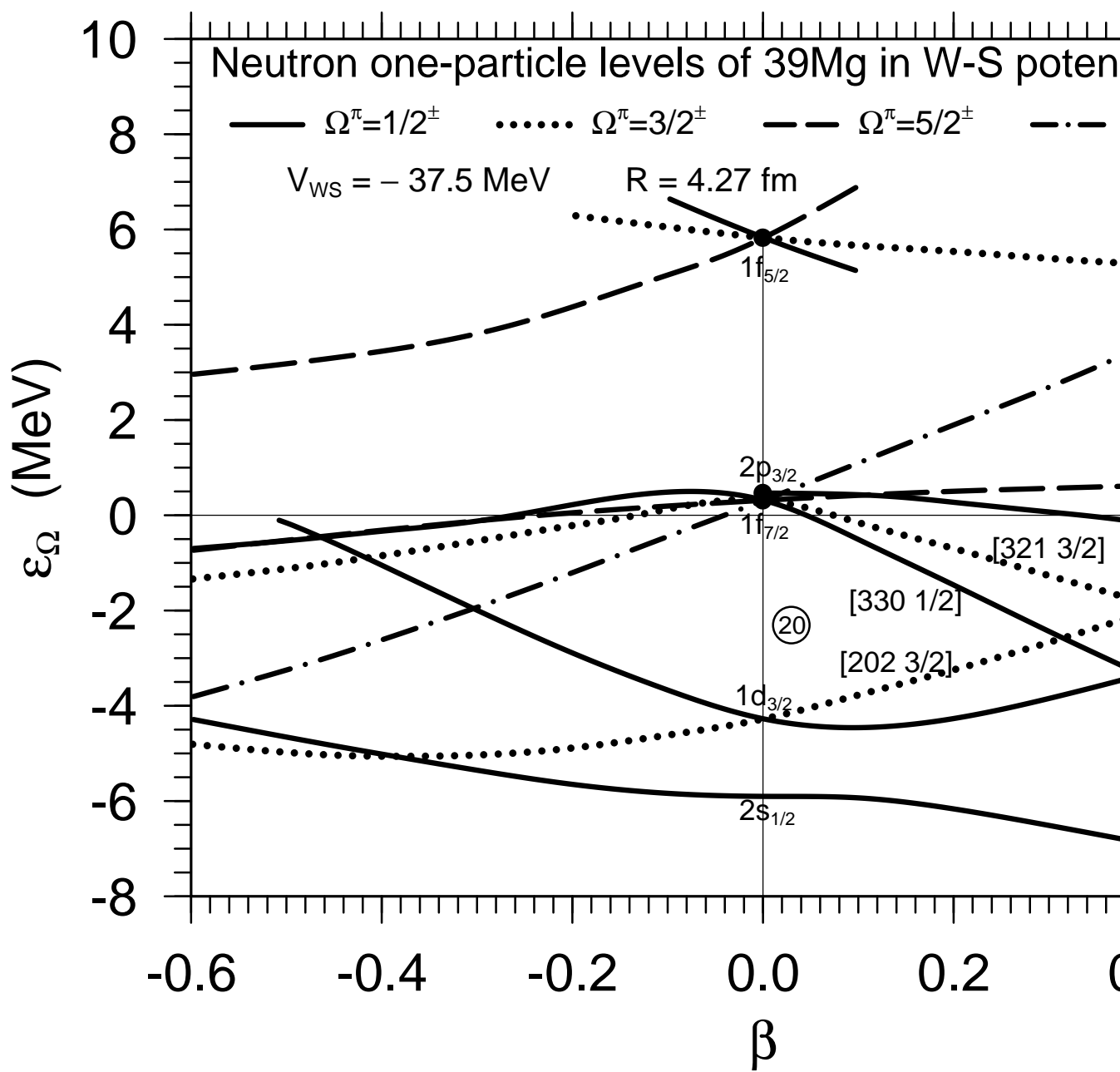
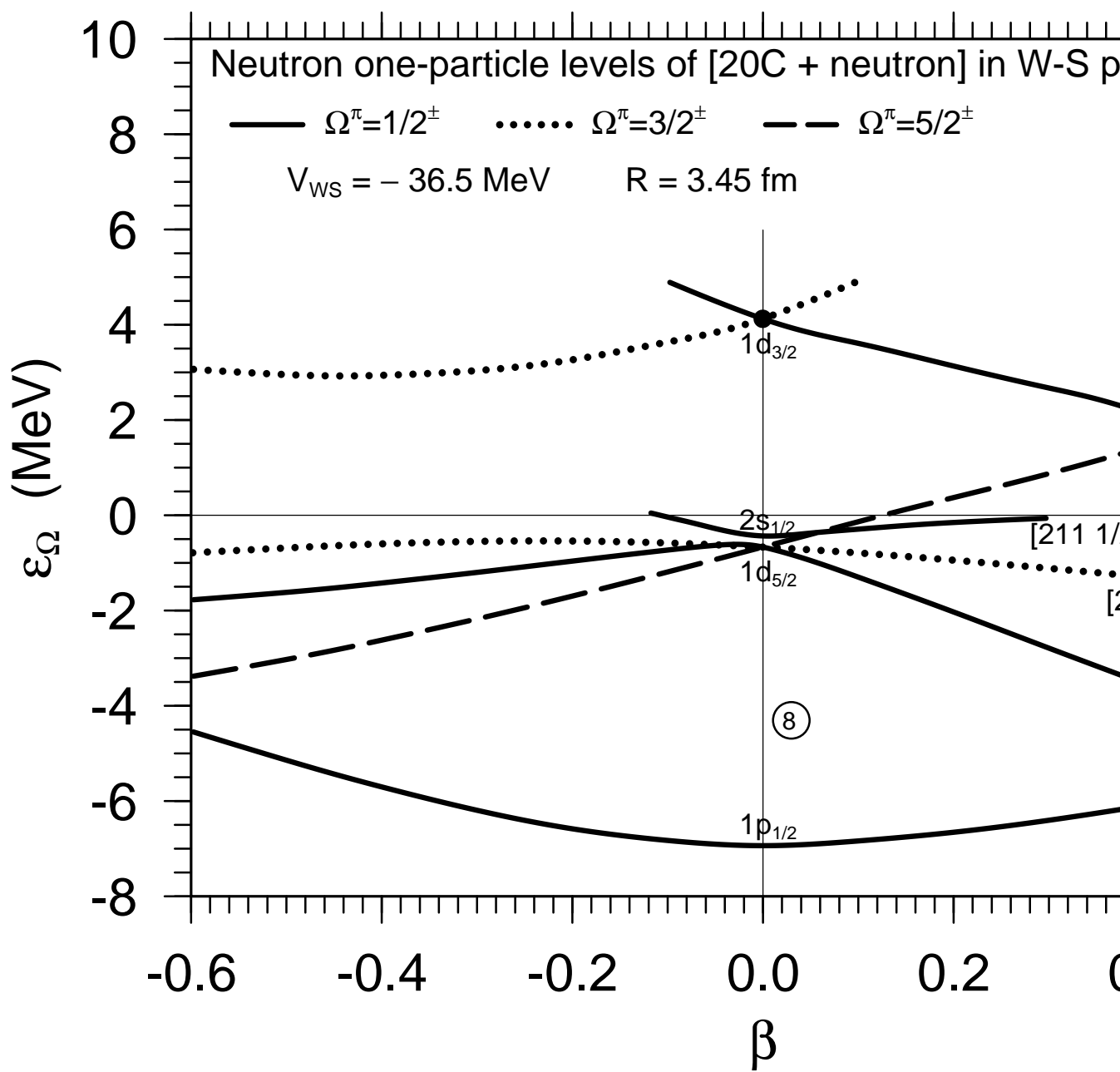
\end{document}